\documentclass[reprint, superscriptaddress, amsmath,amssymb, aps,floatfix]{revtex4-2}

\usepackage{notoccite}
\usepackage{natbib}
\usepackage{graphicx}
\usepackage{dcolumn}
\usepackage{bm}
\usepackage{color,graphicx,subfigure}
\usepackage{enumitem}
\usepackage{wrapfig}
\usepackage{soul}
\usepackage{amsmath}
\usepackage{titlesec}
\usepackage{multirow}
\usepackage{textcomp}
\usepackage{gensymb}
\usepackage[]{hyperref}
\hypersetup{colorlinks=true,linkcolor=blue,citecolor=blue,urlcolor=blue,pdfpagemode=UseNone}
\usepackage{xcolor}
\usepackage{etoolbox}
\usepackage{floatrow}
\usepackage{floatflt}
\usepackage{datetime}
\usepackage{graphicx}
\usepackage{dcolumn}
\usepackage{bm}
\usepackage{lineno}

\definecolor{eh}{rgb}{1, 0, 0}

\definecolor{ks}{rgb}{0.7, 0, 1}

\newcommand{\bscco}{$\textrm{Bi}_2\textrm{Sr}_2\textrm{Ca}\textrm{Cu}_2\textrm{O}_{8+x}$}

\begin{document}
\preprint{APS/123-QED}

\title{Detection of a two-phonon mode in a cuprate superconductor via polarimetric RIXS}

\author{K.\,Scott}
\affiliation{\footnotesize \mbox{Department of Physics, Yale University, New Haven, Connecticut 06520, USA}}
\affiliation{\footnotesize \mbox{Energy Sciences Institute, Yale University, West Haven, Connecticut 06516, USA}}

\author{E.\,Kisiel}
\affiliation{\footnotesize Department of Physics, University of California San Diego, La Jolla, California 92093, USA}

\author{F.\,Yakhou}
\affiliation{\footnotesize European Synchrotron Radiation Facility, 71 Avenue des Martyrs, Grenoble F-38043, France}

\author{S.\,Agrestini}
\affiliation{\footnotesize Diamond Light Source, Harwell Campus, Didcot OX11 0DE, United Kingdom}

\author{M.\,Garcia-Fernandez}
\affiliation{\footnotesize Diamond Light Source, Harwell Campus, Didcot OX11 0DE, United Kingdom}

\author{K.\,Kummer}
\affiliation{\footnotesize European Synchrotron Radiation Facility, 71 Avenue des Martyrs, Grenoble F-38043, France}

\author{J.\,Choi}
\affiliation{\footnotesize Diamond Light Source, Harwell Campus, Didcot OX11 0DE, United Kingdom}

\author{R.\,D.\,Zhong}
\email[Present address: Tsung-Dao Lee Institute and School of Physics and Astronomy, Shanghai Jiao Tong University, Shanghai 200240, China.]{}
\affiliation{\footnotesize Condensed Matter Physics and Materials Science, Brookhaven National Laboratory, Upton, NY, USA}

\author{J.\,A.\,Schneeloch}
\email[Present address: Department of Physics, University of Virginia, Charlottesville, VA 22904, USA]{}
\affiliation{\footnotesize Condensed Matter Physics and Materials Science, Brookhaven National Laboratory, Upton, NY, USA}

\author{G.\,D.\,Gu}
\affiliation{\footnotesize Condensed Matter Physics and Materials Science, Brookhaven National Laboratory, Upton, NY, USA}

\author{Ke-Jin\,Zhou}
\affiliation{\footnotesize Diamond Light Source, Harwell Campus, Didcot OX11 0DE, United Kingdom}

\author{N.\,B.\,Brookes}
\affiliation{\footnotesize European Synchrotron Radiation Facility, 71 Avenue des Martyrs, Grenoble F-38043, France}

\author{A.\,F.\,Kemper}
\affiliation{\footnotesize Department of Physics, North Carolina State University, Raleigh, NC 27695, U.S.A.}

\author{M.\,Minola}
\affiliation{\footnotesize Solid State Spectroscopy Dept, Max Planck Institute for Solid State Research, Stuttgart 70569, Germany}

\author{F.\,Boschini}
\affiliation{\footnotesize Energie Materiaux Telecommunications, Institut National de la Recherche Scientifique, Varennes, Quebec J3X 1S2, Canada}

\author{A.\,Frano}
\affiliation{\footnotesize Department of Physics, University of California San Diego, La Jolla, California 92093, USA}
\affiliation{\footnotesize Canadian Institute for Advanced Research, Toronto, ON, M5G 1M1, Canada}

\author{A.\,Gozar}
\affiliation{\footnotesize \mbox{Department of Physics, Yale University, New Haven, Connecticut 06520, USA}}
\affiliation{\footnotesize \mbox{Energy Sciences Institute, Yale University, West Haven, Connecticut 06516, USA}}

\author{E.\,H.\,da Silva Neto}
\email[Corresponding Author: ]{eduardo.dasilvaneto@yale.edu}
\affiliation{\footnotesize \mbox{Department of Physics, Yale University, New Haven, Connecticut 06520, USA}}
\affiliation{\footnotesize \mbox{Energy Sciences Institute, Yale University, West Haven, Connecticut 06516, USA}}
\affiliation{\footnotesize \mbox{Department of Applied Physics, Yale University, New Haven, Connecticut 06520, USA}}

\begin{abstract}
Recent improvements in the energy resolution of resonant inelastic x-ray scattering experiments (RIXS) at the Cu-\emph{L}$_3$ edge have enabled the study of lattice, spin, and charge excitations. Here, we report on the detection of a low intensity signal at 140\,meV, twice the energy of the bond-stretching (BS) phonon mode, in the cuprate superconductor \bscco~(Bi-2212). Ultra-high resolution polarimetric RIXS measurements allow us to resolve the outgoing polarization of the signal and identify this feature as a two-phonon excitation. Further, we study the connection between the two-phonon mode and the BS one-phonon mode by constructing a joint density of states toy model that reproduces the key features of the data.
\end{abstract}

\maketitle

\section{\label{sec:intro}INTRODUCTION}

In recent years, the resolving power of resonant inelastic x-ray scattering (RIXS) in the soft x-ray regime has improved dramatically to factors of up to 30,000 \cite{ID32}. This gain in resolution has allowed for the observation of novel low-energy features and the disentanglement of known spectral features. Such advancements have enabled RIXS studies of different phenomena in cuprate high-temperature superconductors, including charge order and associated dynamic fluctuations, phonon anomalies, spin excitations, and plasmons \cite{YBCO_spin_excitations,longrange_chargefluc,NCCO_pol,CO_melting,QCDCs_phonon,plasmons,plasmons2,Arpaia_rev}. In the underdoped cuprate \bscco~(Bi-2212), the Cu-\emph{L}$_3$ RIXS energy-loss spectrum below $1$\,eV encompasses an elastic line, a bond-stretching  (BS) phonon at approximately $70$\,meV, and broader features related to damped spin excitations, typically called paramagnons \cite{Le_Tacon_2011}. The improvement in RIXS energy resolution has also enabled the detection of a mode with an energy of approximately $140$\,meV. Although a peak corresponding to this mode is often included when curve-fitting the Bi-2212 RIXS spectrum \cite{bscco_fits}, its origin has not yet been experimentally determined. The energy of this mode falls in a region of interest for Bi-2212 and, as a result, we identified its compatible potential origins to be either excitations across the pseudogap \cite{pseudogap}, or a two-phonon process.

The origin scenarios can be distinguished by resolving the polarization of the scattered x-rays. Most RIXS measurements on the cuprates are performed without outgoing-polarization analysis because of both the scarcity of capable experimental set-ups and the intensive counting times that a polarimetric Cu-\emph{L}$_3$ RIXS experiment requires to achieve the same signal-to-noise ratio (SNR) as a standard RIXS measurement. Here, we report the detection and characterization of the 140\,meV mode via standard RIXS mapping of the $q_x$-$q_y$ scattering plane and a state-of-the-art polarimetric RIXS experiment. The polarimetric measurement reveals that the 140\,meV mode is in the $\sigma$-$\sigma^\prime$ channel, where the un-primed (primed) symbol represents the incident (outgoing) photon polarization. We argue that this result eliminates the possibility the mode originates from a pseudogap, which would be prominent in the $\sigma$-$\pi^\prime$ channel, and indicates a two-phonon process. In the standard RIXS measurements, the mode appears to decrease in intensity with increasing in-plane momentum-transfer magnitude, $q = |\mathbf{q}|$, unlike the BS one-phonon mode which monotonically increases in intensity. In order to understand the contrasting behaviors of these two modes, we model the RIXS intensity of the two-phonon mode as the joint density of states (JDOS) of the BS phonon to capture the behavior of the 140\,meV mode. The polarimetric data, model, and analysis conjointly determine the 140\,meV signal to originate from a two-phonon scattering process.

\section{\label{sec:methods}EXPERIMENTAL DETAILS}

 For all RIXS measurements, the incoming photon energy was set to the Cu-\emph{L}$_3$ absorption peak at $931.5$\,eV and the incoming polarization was set to be in the sample a-b plane (perpendicular to the scattering plane), i.e. incoming $\sigma$-polarization, with the detector positioned at the maximum $2\theta$. Samples were cleaved just prior to their insertion in the ultra-high-vacuum environment of the RIXS chambers. The measurements were performed on two similar underdoped Bi-2212 samples at their respective superconducting transition temperatures ($T_c=60$K and $T_c=54$K). 

\begin{figure}

\includegraphics{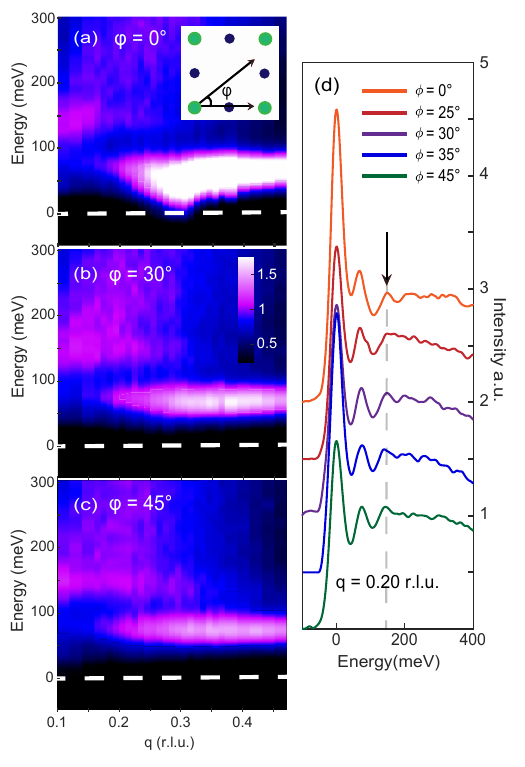}
\caption{\label{fig:qmaps}
Soft X-ray RIXS data at T=T$_{c}$ = 60K for various $\phi$ and $\theta$ orientations in a Bi-2212 sample. (a-c) Energy-loss RIXS spectra mapped along $q$,  at 3 different $\phi$-values. The color bar in (b) applies to (a-c). The elastic line, marked by dashes white lines, has been subtracted from the original data to emphasize the lower-intensity peaks. Raw data is shown in Fig. S1. The inset defines $\phi$ as the angle measured from the Cu-O direction. (d) Stacked individual energy-loss spectra at 5 different $\phi$ orientations at $q$=0.20 r.l.u.. For clarity, the spectra are vertically offset by equal amounts. The arrow and dashed line identify the 140\,meV mode of interest.}

\end{figure}

\emph{Standard RIXS Configuration --} To detect the 140\,meV mode, we performed standard Cu-\emph{L}$_3$ RIXS measurements at Diamond's I21 beamline \cite{ID21}. Standard RIXS measurements, see Fig.\,\ref{fig:qmaps}, were taken at $T=T_c=60$K for $\phi = 0\degree$, $25\degree$, $30\degree$, $35\degree$, and $45\degree$, where $\phi$ is defined as the azimuthal angle from the Cu-O bond direction, see inset of Fig.\,\ref{fig:qmaps}(a), with an energy resolution of $\Delta E=37$meV. The detector was positioned at $2\theta=154\degree$. For each $\phi$, we acquired RIXS spectra at different values of $q$ by varying the incident angle, $\theta$, on the sample. Values of $q$ are reported in reciprocal lattice units (r.l.u.), where one r.l.u. is defined as $2\pi/a$ and $a = 3.82$\,\AA~(the distance between nearest neighbor Cu atoms). 
The standard (non-polarimetric) RIXS data used here is the same data set as used in Ref.\,\cite{QCDCs_phonon} to map the energy-momentum structure of the BS phonon ($E\leq72$\,meV) and to detect quasi-circular dynamic correlations associated with charge order in Bi-2212. In the present report, we analyze a different feature of this data, the signal at $E\approx140$\,meV.

\begin{figure*}
\includegraphics{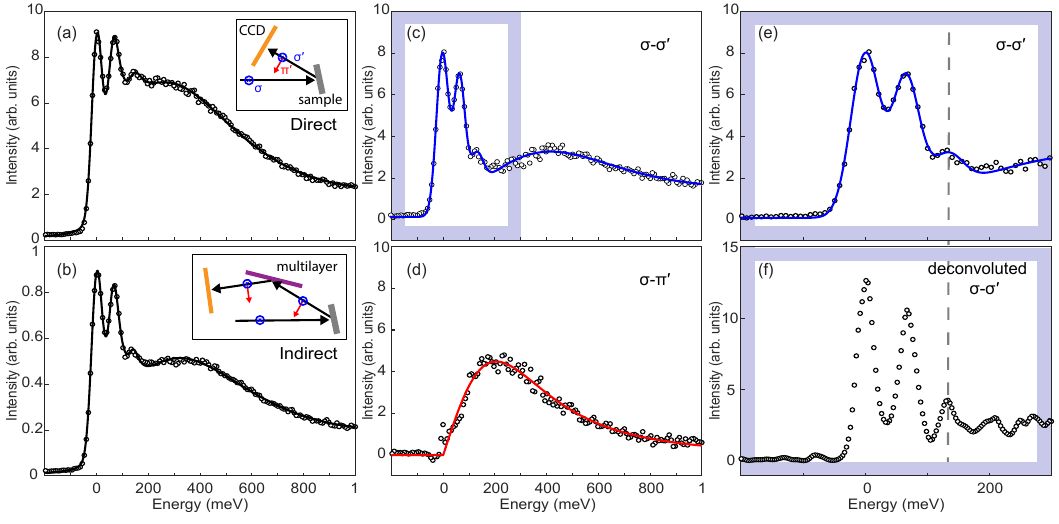}
\caption{\label{fig:pol} Polarimetric RIXS data at $q$=0.25 r.l.u., $\phi=45\degree$, T=54K, $\Delta E=41$meV. (a, b) The raw spectra as measured on the direct (a) and indirect (b) detectors. The black lines are generated with the fitting procedure shown in Fig.\,\ref{fig:fits}(a) and detailed in the Supplemental Material. The insets of (a) and (b) depict a schematic of the scattering plane (orthogonal to the sample a-b plane) for their respective experimental set-ups . The blue (red) vectors indicate $\sigma$- ($\pi$-) polarized light. The orange represents the CCD detector and the purple represents the multilayer (ML) mirror. The geometry of the insets reflect the experiment ($2\theta$=149.5\degree, $\theta$=102\degree). (c, d) The spectra decomposed into $\sigma$-$\sigma^\prime$ (c) and $\sigma$-$\pi^\prime$ (d) polarization. The blue line (c) is generated with the same fitting procedure, while the red (d) uses only a paramagnon. (e) A zoomed-in depiction of the data in (c) with the gray box indicating the region. (f) The first three peaks in the $\sigma$-$\sigma^\prime$ spectra, calculated using deconvoluted direct and indirect spectra (see text for details).}
\end{figure*}

\emph{Polarimetric RIXS Configuration --} To resolve the mode's outgoing polarization, we performed polarimetric Cu-\emph{L}$_3$ RIXS measurements at ESRF's ID32 beamline \cite{ID32}, the only beamline in the world capable of resolving outgoing polarization in the soft x-ray regime. The polarimetric RIXS measurements (Fig.\,\ref{fig:pol}) were taken at $\phi=45\degree$, $q=0.25$ r.l.u., $T=T_c=54$K, with an energy resolution of $\Delta E=41$meV.  The detector was positioned at $2\theta=149.5\degree$. In this mode of operation, two consecutive spectra are taken with and then without a multilayer mirror to resolve the polarization of the scattered photons \cite{Braicovich}. In the direct measurement (standard RIXS configuration), shown in the inset of Fig.\,\ref{fig:pol}(a), incoming $\sigma$-polarized x-rays scatter from the sample. The scattered outgoing light, composed of mixed $\sigma^\prime$ and $\pi^\prime$-polarized rays, is collected on a charge-coupled device (CCD) detector. In the indirect, polarimetric measurement, shown in the inset of Fig.\,\ref{fig:pol}(b), the outgoing light reflects from a multilayer mirror with different reflectivities for $\sigma^\prime$- and $\pi^\prime$-polarized light before it is collected on a second CCD detector. Due to the low reflectivities of the mirror (R$_{\sigma} \approx$ 0.1 and R$_{\pi} \approx$ 0.05 for $\sigma^\prime$- and $\pi^\prime$-polarized light, respectively), the indirect signal measurement requires a ten-fold increase in counting times to maintain a reasonable SNR. In previous polarimetric RIXS measurements, the acquisition time was often shortened by compromising on energy resolution ($\Delta E \sim 90$meV) to increase flux \cite{NBCO_pol,NCCO_pol}. However, ultra-high energy resolution ($\Delta E \sim 40$\;meV) is required to detect the relatively lower intensity $140$\,meV mode. The direct and indirect spectral intensities are given by
\begin{equation}
    I_{direct}=I_{\sigma^\prime}+I_{\pi^\prime}
    \label{Idir}
\end{equation}
\begin{equation}
    I_{indirect}=R_{\sigma^\prime} I_{\sigma^\prime} + R_{\pi^\prime} I_{\pi^\prime}
    \label{Iindir}
\end{equation}
From here, the outgoing $\sigma^\prime$- and $\pi^\prime$-polarized spectra can be calculated, using
\begin{equation}
 I_{\pi^\prime} = \frac{I_{indirect} - R_{\sigma^\prime} I_{direct}}{R_{\pi^\prime}-R_{\sigma^\prime}}
 \label{Ipi}
\end{equation}
\begin{equation}
 I_{\sigma^\prime} = \frac{I_{indirect} - R_{\pi^\prime} I_{direct}}{R_{\sigma^\prime}-R_{\pi^\prime}}
 \label{Isig}
\end{equation}

\section{\label{sec:results}RESULTS}

Figures \ref{fig:qmaps}(a-c) shows RIXS energy-momentum maps for three different azimuthal orientations. To highlight the lower intensity features, we show the spectra, in the range between -50 and 300 meV, after subtraction of the elastic line. In this energy range, these spectra feature the well-known BS phonon at approximately 70\,meV and broad spin excitations stretching across the entire energy range. A third feature is also visible, appearing as a soft streak of intensity at roughly constant 140\,meV energy, observed in all measured $\phi$-orientations. The RIXS spectra at $q$ = 0.20 r.l.u. for different $\phi$ orientations are shown in Fig.\,\ref{fig:qmaps}(d), where the 140 meV feature is clearly visible. The spectral intensity in this energy region has no apparent $\phi$-dependence, but does decrease with higher $q$. However, without outgoing polarization resolution, no definitive conclusions can be made regarding the origin of the mode.

\begin{figure*}
\includegraphics{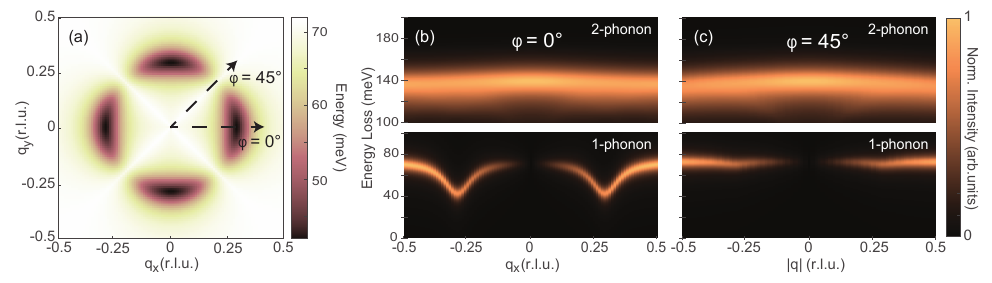}
\caption{\label{fig:toymodel} Toy JDOS model of one-phonon and two-phonon dispersions. (a) Dispersion of the single phonon in the q$_x$-q$_y$ plane. The black dashed lines define the $\phi$-orientations of panels b and c. (b-c) Calculated RIXS intensities of the single phonon and two-phonon modes against $q$ along the two high-symmetry directions as predicted by the model. The modes are individually normalized by their maximum for comparison.}
\end{figure*}

Given the energy scale of the mode, we hypothesize two possible origin scenarios: a peak related to pseudogap,  or a two-phonon process. First, below 150K, Bi-2212 has well-known pseudogap with an energy of 140\,meV across the gap \cite{pseudogap}. Second, 140\,meV is approximately twice the energy of the BS phonon, which is weakly dispersing in the absence of charge order. In Raman spectroscopy, the pseudogap feature appears in the cross-polarized B$_{1g}$ channel \cite{pseudogap}. On the other hand, Raman measurements typically find two-phonon modes to be fully symmetric, appearing in the non-crossed polarized $A_{1g}$ channel \cite{twophonon_transitionmetals_Raman, Cardona1983}. While these Raman measurements probe excitations at $q$=0, i.e. the Brillouin zone center, we expect the relative symmetries to hold at finite $q$. In other words, although the B$_{1g}$ symmetry of a signal at $q$=0 is not preserved at finite $q$, the signal remains cross-polarized. In a RIXS measurement with incoming $\sigma$-geometry, a B$_{1g}$ pseudogap signal would be prominent in scattering $\phi = 45\degree$ away from the Cu-O bond direction and in the cross-polarized $\sigma$-$\pi^\prime$ channel \cite{Raman_geometry}. A two-phonon process would be detectable for all $\phi$-orientations in the non-crossed polarized $\sigma$-$\sigma^\prime$ channel. We distinguish these scenarios by resolving the polarization of the outgoing light  with ultra-high energy resolution in a polarimetric RIXS experiment.

In Figs.\,\ref{fig:pol}(a) and \ref{fig:pol}(b), the indirect and direct spectra below 1eV are shown, each encompassing the expected features (an elastic line, BS phonon, the 140\,meV mode, and spin excitations) with differing relative intensities. The spin excitations are known to contain both a lower-energy component related to a single spin flip and a higher-energy component related to a double spin flip \cite{Betto_PRB_2021_Multiple_magnon}. Applying Eqs. \ref{Ipi} and \ref{Isig} to the spectra yields the $\sigma$-$\pi^\prime$ and $\sigma$-$\sigma^\prime$ spectra shown in Figs.\,\ref{fig:pol}(c) and \ref{fig:pol}(d). As expected, the $\sigma$-$\pi^\prime$ channel contains the lower-energy single-flip spin excitations. The elastic line, BS phonon and higher-energy double-flip spin excitations are prominent in the $\sigma$-$\sigma^\prime$ channel. Notably, the 140\;meV mode is present in the $\sigma$-$\sigma^\prime$ channel, as more clearly observed by zooming in on the $\sigma$-$\sigma^\prime$ spectrum, Fig.\,\ref{fig:pol}(e). Recalculating the $\sigma$-$\sigma^\prime$ spectra from the deconvoluted direct and indirect spectra in Fig.\,\ref{fig:pol}(f) further emphasizes that mode. [See supplemental materials for details of the deconvolution procedure \footnote{See Supplemental Material at [URL will be inserted by publisher] for additional RIXS data and analysis, details on the JDOS model, error propagation, and polarimetric RIXS parameters, which includes Refs. \cite{yang_Deconv_2008,para_fit,barevertex,QCDCs_phonon}.}.] These results rule out the pseudogap scenario and demonstrate that the signal is generated from a two-phonon process.

\section{\label{sec:analysis} DISCUSSION}

The two-phonon mode exhibits two distinct behaviors whose connection is not initially intuitive. First, while the one-phonon mode exhibits a softening in energy around q$_{CO}=0.27$ r.l.u., which is most prominent along the $\phi=0$ direction \cite{CO_melting, QCDCs_phonon}, the two-phonon mode is near-constant in energy, as shown in Fig.\,\ref{fig:qmaps}. Second, while the total spectral intensity in the energy range of the one-phonon mode increases at higher $q$, the intensity in the energy range of the two-phonon mode decreases at higher q. To understand the relationship between the one-phonon and two-phonon modes, we construct a two-phonon toy model. The model emulates two-phonon mode behavior by modeling the signal as the JDOS of two one-phonon scatterings and implementing the appropriate kinematic constraints \cite{RIXS_phonons}. Specifically, the positions in $\mathbf{q}$-space and energy of the two-phonon are limited to the possible summations of $\mathbf{q}$ and energy from two one-phonons: $\mathbf{q}_{2p}=\mathbf{q}_{1}+\mathbf{q}_{2}$, $\omega_{2p}=\omega_1+\omega_2$. The model starts from the one-phonon dispersion, $\epsilon(\mathbf{q})$, in the q$_x$-q$_y$ plane, Fig.\,\ref{fig:toymodel}(a), which is constructed from our experimental data \footnotemark[\value{footnote}]. This dispersion exhibits the softening of the BS phonon around $q\approx q_{CO}$ in the $\phi =0 \degree$ direction and the monotonic trend in energy of the BS phonon along the $\phi =45 \degree$ direction. Using this dispersion, we calculate the density of states of the BS phonon using the following Green's function

\begin{equation}
    D(\omega, \mathbf{q}) = \frac{1}{\omega-\epsilon(\mathbf{q})+i\Gamma}
    \label{greens}
\end{equation}
Then, to model the scattering cross-section we assume the bare vertex electron-phonon coupling form \cite{barevertex} 
\begin{equation}
    g(q_x,q_y)\propto \sqrt{\sin^2{(q_x \* \pi)}+\sin^2{(q_y \* \pi)}}
    \label{e-ph}
\end{equation}
where q$_x$ and q$_y$ are represented in r.l.u..

The results in Figs.\,\ref{fig:toymodel}(b) and \ref{fig:toymodel}(c) show that, in contrast to the behaviors of the inputted one-phonon mode, the model predicts a two-phonon mode that is near-constant in energy for both high-symmetry directions. Thus, the model explains the lack of softening of the two-phonon mode in the charge order regime.

\begin{figure}
\includegraphics[width=8cm]{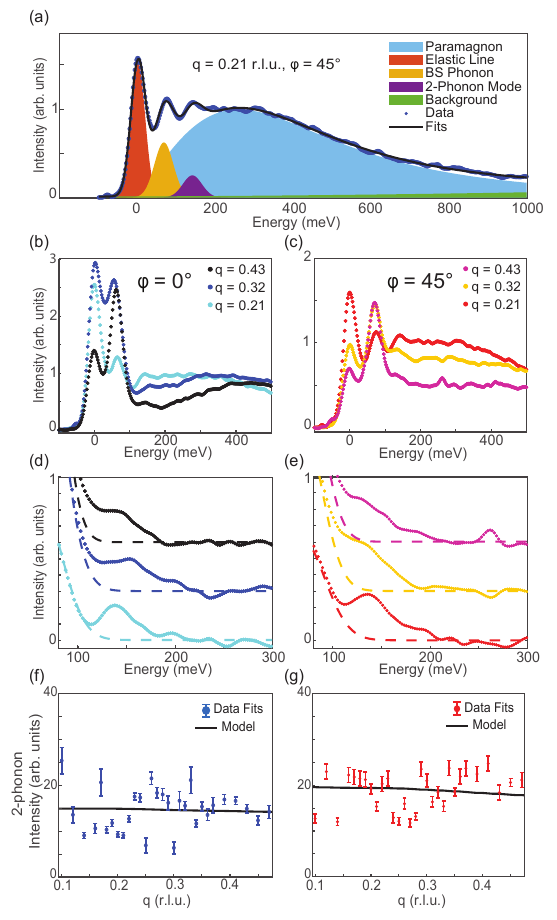}
\caption{ Fitting the 140 meV mode. (a) An example of the fitting procedure. The dark blue line is the composite fit to the spectra. It includes 5 components: an elastic peak at 0 meV (orange), the BS $\sim$70 meV phonon peak (yellow), the novel 140 meV peak (purple), a paramagnon continuum (light blue) and a broad background (green). (b-c) The raw spectra show the relative intensities at 140\,meV along the two high symmetry directions at 3 different $q$-values. (d-e) The spectra shows the intensity of 140 meV mode after subtracting the paramagnon and background contributions. The dashed line indicates the tail of the BS phonon peak. (f-g) Comparison of the model and measurements. The blue (red) markers are the integrated intensities of the fit to the 140 meV peak as shown in (a) in the $\phi$=0\degree (45\degree) direction. The error bars were propagated using the 95$\%$ confidence intervals on the fits, as outlined in the Supplemental Material. The black line is the predicted intensity (integrated from 120\,meV to 150\,meV) of the JDOS model, normalized to the average intensity of the integrated fits.}
\label{fig:fits}
\end{figure}

To compare the predicted flat two-phonon intensity to the measured two-phonon intensity, we implement a curve fitting analysis to experimental data. The fitting analysis shown in Fig.\,\ref{fig:fits}(a) includes an elastic line, a one-phonon peak, a two-phonon peak, a paramagnon, and a broad background. In the fitting function \footnotemark[\value{footnote}], the elastic line, one-phonon peak, and two-phonon peak are modelled as Gaussians, the paramagnon as an anti-symmetric Lorentzian \cite{para_fit}, and the background as a second-order polynomial. The two-phonon mode overlaps in energy with the paramagnon and, since the paramagnon shifts to higher energy with increasing $q$, the total signal intensity in the 140\,meV region of the non-dispersive two-phonon peak decreases, as shown in Figs.\,\ref{fig:fits}(b) and \ref{fig:fits}(c), for both high symmetry directions. However, subtracting the fitted paramagnon and background features from the spectra reveals a two-phonon mode that is near-constant in intensity, Figs.\,\ref{fig:fits}(d) and \ref{fig:fits}(e). Similarly, directly implementing the fit to the two-phonon mode yields a flat integrated peak intensity as a function of $q$, as shown in Figs.\,\ref{fig:fits}(f) and\ref{fig:fits}(g). [This $q$-dependence also confirms the prominence of the mode across the Brillouin zone.] Next, we integrate the two-phonon intensity predicted by the JDOS model (black line) and normalize it for comparison to the data. Despite the low SNR of the extracted two-phonon peak, the intensity trends of the model and experiment are in agreement. Thus, employing the JDOS model in tandem with the fitting procedure resolves the perceived discrepancy between the intensities of the BS phonon signal and the two-phonon signal. 
This intensity analysis also proves incompatible with a previously considered origin: the proposed B$_{2g}$-channel charge order gap, detected in  Hg- and Y-based cuprates via Raman spectroscopy \cite{COgap}. In the charge order gap interpretation, one expects a spectral feature with a significant change in the energy-momentum structure in the vicinity of $q \approx q_{CO}$ \cite{COgap_theory}, which clearly disagrees with the flat intensity of our experimental data. Therefore, the intensity profile of the 140\,meV mode robustly evidences a two-phonon process.

\section{\label{sec:conclusion} CONCLUSION}

Often the first approach to resolve subtle RIXS signals is to improve energy resolution. However, decomposition of the scattering processes into different polarimetric channels has the practical effect of separating overlapping signals, making it an important tool in situations where energy-resolution cannot be improved.
We leveraged this capability in our work, where we performed a polarimetric RIXS experiment, aided by standard RIXS measurements and model calculations, to identify the origin of the subtle 140\,meV mode. Our data demonstrated that the 140\,meV mode is a two-phonon mode: its behavior in intensity and energy as a function of $q$ is consistent with two-phonon scattering, and it appears in the $\sigma$-$\sigma^\prime$ channel. Polarimetric decomposition with high energy-resolution provides a significant advance in soft x-ray RIXS, but it requires significantly longer counting times.
Previous polarimetric \mbox{Cu-\emph{L}$_3$} RIXS measurements have focused on features that are more intense and broader in energy than the two-phonon peak, such as paramagnon, magnon and bi-magnon excitations, where energy-resolution could be compromised in favor of a larger scattered beam flux.
On the other hand, the polarimetric resolution of the two-phonon feature existing amongst much more intense magnetic and single-phonon peaks required the highest-possible energy resolution. As such, our experiments sets a new state-of-the-art standard for polarimetric soft x-ray RIXS measurements.
 
\section*{ACKNOWLEDGEMENTS} We acknowledge the Diamond Light Source, UK, for time on beamline I21-RIXS under proposals MM28523 and MM30146. We acknowledge the European Synchrotron Radiation Facility, France, for time on beamline ID32 for polarimetric RIXS under proposal HC-4824. This material is based upon work supported by the National Science Foundation under Grant No. DMR-2034345 and DMR-2145080. The work at BNL was supported by the US Department of Energy, office of Basic Energy Sciences, contract no. DOE-sc0012704. A.F.K. was supported by the NSF under grant no. DMR-1752713. A.F. was supported by the Research Corporation for Science Advancement via the Cottrell Scholar Award (27551) and the CIFAR Azrieli Global Scholars program. This work was supported by the Alfred P. Sloan Fellowship (E.H.d.S.N.). F.B. acknowledges support from the Fonds de recherche du Quebec – Nature et technologies (FRQNT) and the Natural Sciences and
Engineering Research Council of Canada (NSERC).

\bibliography{apssamp}

\end{document}


\title{\textit{Supplementary Materials}: Detection of a two-phonon mode in a cuprate superconductor via polarimetric RIXS}

\author{K.\,Scott}
\affiliation{\footnotesize \mbox{Department of Physics, Yale University, New Haven, Connecticut 06520, USA}}
\affiliation{\footnotesize \mbox{Energy Sciences Institute, Yale University, West Haven, Connecticut 06516, USA}}

\author{E.\,Kisiel}
\affiliation{\footnotesize Department of Physics, University of California San Diego, La Jolla, California 92093, USA}

\author{F.\,Yakhou}
\affiliation{\footnotesize European Synchrotron Radiation Facility, 71 Avenue des Martyrs, Grenoble F-38043, France}

\author{S.\,Agrestini}
\affiliation{\footnotesize Diamond Light Source, Harwell Campus, Didcot OX11 0DE, United Kingdom}

\author{M.\,Garcia-Fernandez}
\affiliation{\footnotesize Diamond Light Source, Harwell Campus, Didcot OX11 0DE, United Kingdom}

\author{K.\,Kummer}
\affiliation{\footnotesize European Synchrotron Radiation Facility, 71 Avenue des Martyrs, Grenoble F-38043, France}

\author{J.\,Choi}
\affiliation{\footnotesize Diamond Light Source, Harwell Campus, Didcot OX11 0DE, United Kingdom}

\author{R.\,D.\,Zhong}
\affiliation{\footnotesize Condensed Matter Physics and Materials Science, Brookhaven National Laboratory, Upton, NY, USA}

\author{J.\,A.\,Schneeloch}
\affiliation{\footnotesize Condensed Matter Physics and Materials Science, Brookhaven National Laboratory, Upton, NY, USA}

\author{G.\,D.\,Gu}
\affiliation{\footnotesize Condensed Matter Physics and Materials Science, Brookhaven National Laboratory, Upton, NY, USA}

\author{Ke-Jin\,Zhou}
\affiliation{\footnotesize Diamond Light Source, Harwell Campus, Didcot OX11 0DE, United Kingdom}

\author{N.\,B.\,Brookes}
\affiliation{\footnotesize European Synchrotron Radiation Facility, 71 Avenue des Martyrs, Grenoble F-38043, France}

\author{A.\,F.\,Kemper}
\affiliation{\footnotesize Department of Physics, North Carolina State University, Raleigh, NC 27695, U.S.A.}

\author{M.\,Minola}
\affiliation{\footnotesize Solid State Spectroscopy Dept, Max Planck Institute for Solid State Research, Stuttgart 70569, Germany}

\author{F.\,Boschini}
\affiliation{\footnotesize Energie Materiaux Telecommunications, Institut National de la Recherche Scientifique, Varennes, Quebec J3X 1S2, Canada}

\author{A.\,Frano}
\affiliation{\footnotesize Department of Physics, University of California San Diego, La Jolla, California 92093, USA}
\affiliation{\footnotesize Canadian Institute for Advanced Research, Toronto, ON, M5G 1M1, Canada}

\author{A.\,Gozar}
\affiliation{\footnotesize \mbox{Department of Physics, Yale University, New Haven, Connecticut 06520, USA}}
\affiliation{\footnotesize \mbox{Energy Sciences Institute, Yale University, West Haven, Connecticut 06516, USA}}

\author{E.\,H.\,da Silva Neto}
\affiliation{\footnotesize \mbox{Department of Physics, Yale University, New Haven, Connecticut 06520, USA}}
\affiliation{\footnotesize \mbox{Energy Sciences Institute, Yale University, West Haven, Connecticut 06516, USA}}
\affiliation{\footnotesize \mbox{Department of Applied Physics, Yale University, New Haven, Connecticut 06520, USA}}

\maketitle

\noindent\textbf{I. Polarimetric analysis parameters}\\
The direct and indirect spectra were collected for over 72 hours, with the indirect-to-direct counting time ratio equal to 12, to ensure sufficient statistics for the polarimetric decomposition. The direct (indirect) spectra was averaged from $\sim$2000 ($\sim$4000) individual spectra with 10s (60s) exposures, which were carefully aligned and summed. We calculated the reflectivities of the multilayer for each incoming polarization ($R_\pi=0.05211$ and $R_\sigma=0.09728$) by measuring the elastic line of silver paint with and without the inclusion of the multilayer, fitting a Gaussian peak to those signals to calculate the integrated intensity, and dividing the quantities. The errors calculated the errors $\sigma_\pi=0.0032$ and $\sigma_\sigma=0.0074$.
To ensure that 1) there is no negative signal intensity in either channel and 2) the elastic line and BS phonon both appear in the $\sigma$-$\sigma^\prime$ channel, we used $R_\pi=0.05211$ and $R_\sigma=0.10468$, which are within the bounds of error of the values obtained on silver paint.

\noindent\textbf{II. Deconvolution procedure}\\
The Lucy-Richardson deconvolution procedure \cite{yang_Deconv_2008} was used to deconvolve the energy resolution ($\sim$41\,meV) from the direct and indirect RIXS spectra in Fig. 2 of the main text. Deconvolution parameters (number of iterations, region of interest) were optimized by ensuring that the reconvolution of the deconvolved curves reproduced the original data. A comparison between the raw and deconvolved spectra is shown in Fig. \ref{fig:deconv_specs}.

\noindent\textbf{III. Fitting procedure}\\ 
To analyze the two-phonon mode intensity, we implemented the following five-peak fit function to the standard RIXS spectra

\begin{equation}
        f(\omega) =   \sum_{i=1}^{3} A_{i} e^{-\bigl(\frac{\omega-\omega_{i}}{c_{i}}\bigr)^2}+ 
    \frac{1}{1+e^{-2k\omega}}\cdot \frac{A \gamma \omega}{\pi((\omega^2-\omega_0^2)^2+4\omega^2\gamma^2))}+
    \sum_{j=0}^{2}a_j\omega^j
    \label{eq:fits}
\end{equation}
in which the three Gaussian peaks represent the elastic line, BS phonon mode, and two-phonon mode with amplitude $A_{i}$ position $\omega_{i}$ and width $c_{i}$. The paramagnon is represented by the heavy-side function (k=10$^4$\,meV$^{-1}$ for a sharp drop at $\omega$=0) multiplied by an anti-Lorentzian with amplitude A, position $\omega_{0}$ and width $\gamma$ \cite{para_fit}. The smooth background is represented by a parabola with constants given by $a_j$. The width parameter is constrained below 45\,meV (consistent with the energy resolution $\Delta E = 37$\,meV) for the Gaussian peaks. Due to its overlap with the paramagnon feature, the two-phonon peak position was constrained to 125\,meV $< \omega_{2p}<$ 145\,meV. For the spectra with lowest BS phonon intensity (corresponding to q $<$ 0.17 r.l.u.), the BS phonon peak position was constrained to $\omega_{1p}>$ 61\,meV. All other parameters were left free.

\noindent\textbf{IV. Toy model calculation}\\
In this phenomenological model, the two-phonon spectral intensity is derived from the joint density of states (JDOS) of two one-phonons, which enforces the kinematic contraints: $\mathbf{q}_{2p}=\mathbf{q}_{1}+\mathbf{q}_{2}$ and $\omega_{2p}=\omega_1+\omega_2$. 
The one-phonon dispersion is modeled as
\begin{equation}
    \epsilon(\mathbf{q}) = \epsilon_0 - \xi(\phi) \frac{\Delta}{(\frac{q-q_0}{\gamma_q})^2 + 1}
\end{equation}
where $q=|\mathbf{q}|$ and the parameters $\epsilon_0 = 72$\,meV, $\Delta = 30$\,meV, $\gamma_q = 0.065$\,r.l.u., $q_0 = 0.29$\,r.l.u.~and $\xi(\phi) = (|\cos(2\phi)| + 0.08)/(1.08)$ are adjusted to match the experimental data from Fig. 1 in the main text, as in Ref. \cite{QCDCs_phonon}. The density of states of two one-phonons can each be modeled independently via the a Green's functions
\begin{equation}
    D(\omega,\mathbf{q})=\frac{1}{\omega-\epsilon(\mathbf{q})+i\Gamma}
\end{equation}
where $\Gamma$=50\,meV. The imaginary part of which gives the spectral weight of the phonon. Thus, we invoke the bare-vertex form of the electron-phonon coupling \cite{barevertex}
\begin{equation}
    g(\mathbf{q})=g_0A_{Cu}A_O\sqrt{\sin^2{(\pi q_x)} + \sin^2{(\pi q_y)}}
\end{equation}
where we set $g_0A_{Cu}A_O\equiv1$, and obtain the weighted spectral function
\begin{equation}
    A(\omega,\mathbf{q})=-\mathrm{Im}[D(\omega,\mathbf{q})] \cdot g(\mathbf{q})
\end{equation}
The spectral intensity of the two-phonon mode can then be described by the convolution of two one-phonon spectral functions
\begin{equation}
    J(\mathbf{q},\omega)=\int dk^2 d\omega^\prime [A(\mathbf{k},\omega^\prime) A(\mathbf{k-q},\omega-\omega^\prime) ]
\end{equation}
which we calculated following the convolution theorem using Fourier transforms
\begin{equation}
 J(\mathbf{q},\omega) = \mathcal{F}^{-1}[\mathcal{F}(A(\mathbf{q},\omega)) \cdot\mathcal{F}(A(\mathbf{q},\omega))]
\end{equation}

 \noindent\textbf{V. Error propagation}\\

The errors on the multilayer reflectivities were calculated using
 \begin{equation}
     \sigma_R = R \cdot \sqrt{\biggl(\frac{\sigma_{direct}}{I_{direct}}\biggr)^2 + \biggl(\frac{\sigma_{indirect}}{I_{indirect}}\biggr)^2}
 \end{equation}
 where R is the reflectivity for a given incoming polarization, $I_{direct}$ ($I_{indirect}$) is the integrated intensity of the fit to the silver paint measurement without (with) the multilayer reflection and $\sigma_{direct}$ ($\sigma_{indirect}$) is the uncertainly to that fit, as determined by the RixsToolBox processing software.

 The error bars on the two-phonon integrated intensity were propagated using
 \begin{equation}
     \sigma_3 = I_3 \cdot \sqrt{\biggl(\frac{\sigma_{A_3}}{A_3}\biggr)^2 + \biggl(\frac{\sigma_{c_3}}{c_3}\biggr)^2}
 \end{equation}
 where $A_3$ and $c_3$ are the fit parameters and $I_3$ is the integration of the Gaussian fit to the two-phonon in Eq. \ref{eq:fits}. The errors to these parameters, $\sigma_{A_3}$ and $\sigma_{c_3}$, were extracted from the MATLAB-generated 95\% confidence intervals. 

\noindent\textbf{VI. Standard RIXS maps, before elastic line subtraction}

\begin{figure}[H]
    \centering
    \includegraphics[width=16cm]{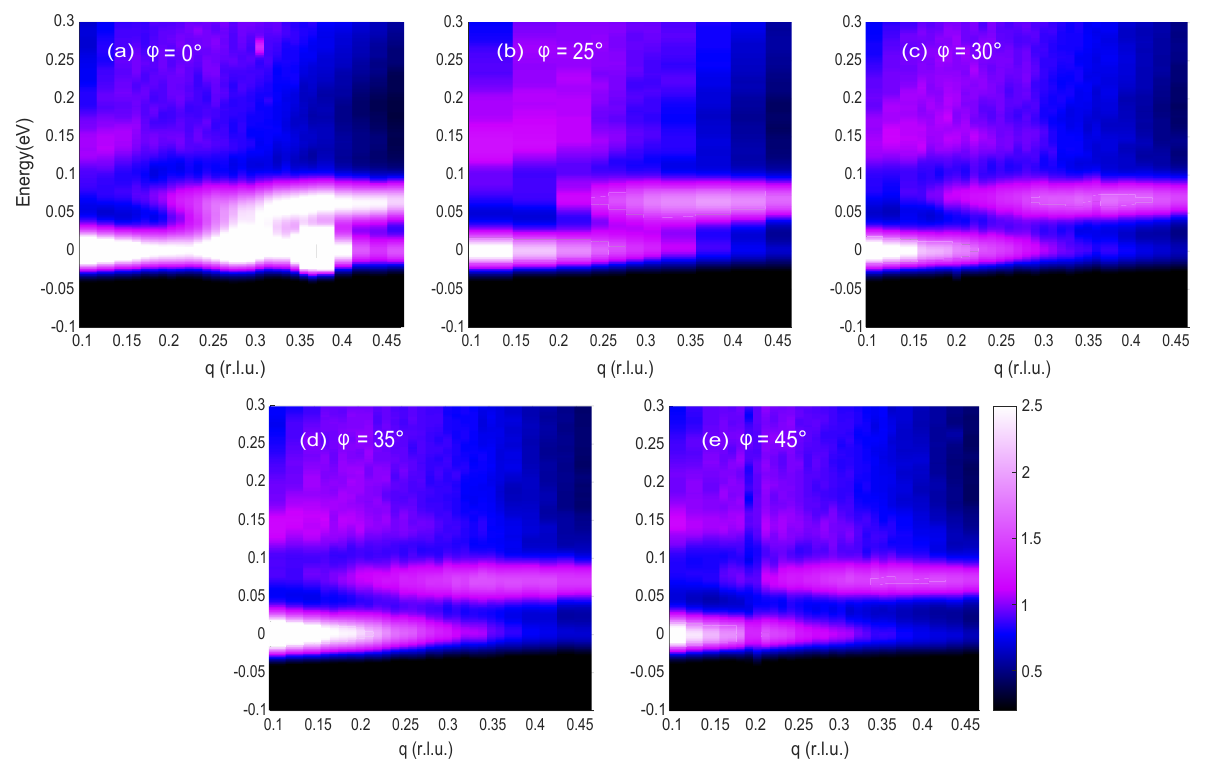}
    \caption{Soft X-ray RIXS data at T=T$_{c}$ = 60K for various $\phi$ and $\theta$ orientations in a Bi-2212 sample. RIXS spectra for $\phi$ = 0\degree, 25\degree, 30\degree, 35\degree, 45\degree\, before subtraction of elastic line. The maps have been normalized by their total respective intensity and shifted to ensure zero intensity in the E$<$0 range. The zero- energy points were aligned using the fits to the elastic line in \ref{fig:EL fits}. The discontinuous spectra in (e) at q=0.19 and 0.20 were interpolated over in the main text analysis.}
    \label{fig:maps with EL}
\end{figure}

\newpage
\noindent\textbf{VII. Fits to elastic lines}

\begin{figure}[H]
    \centering
    \includegraphics[width=16cm]{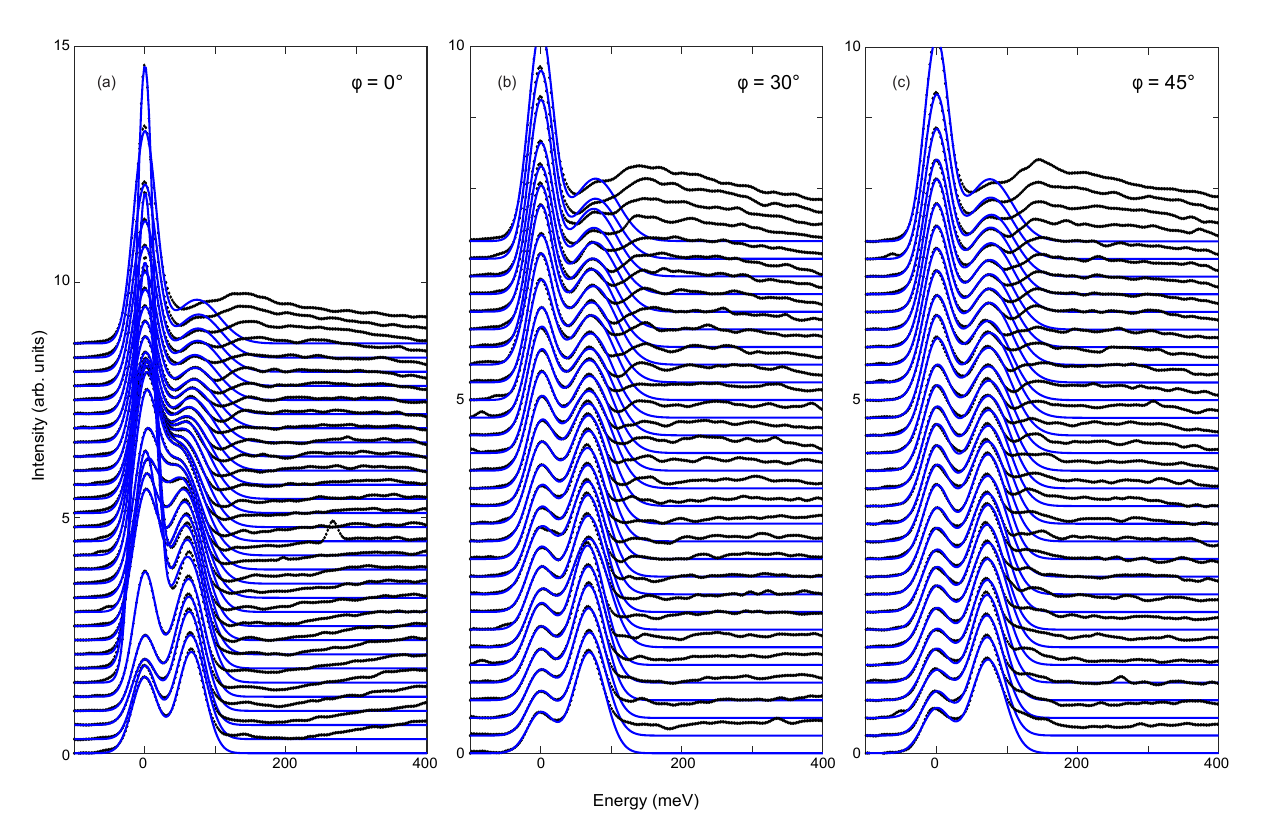}
    \caption{Simple Gaussian fits to the first two peaks in the $<$ 100 meV region (elastic line and BS phonon) for the three $\phi$-orientations in Fig. 1 of the main text, used for subtracting the elastic line for emphasis on lower intensity figures. The spectra correspond (top to bottom) to $q$-values (in r.l.u.): 0.10 to 0.16 in steps of 0.02, 0.17 to 0.37 in steps of 0.01, 0.39 to 0.47 in steps of 0.02.}
    \label{fig:EL fits}
\end{figure}

\noindent\textbf{VIII. Deconvoluted Spectra}

\begin{figure}[H]
    \centering
    \includegraphics[width=16cm]{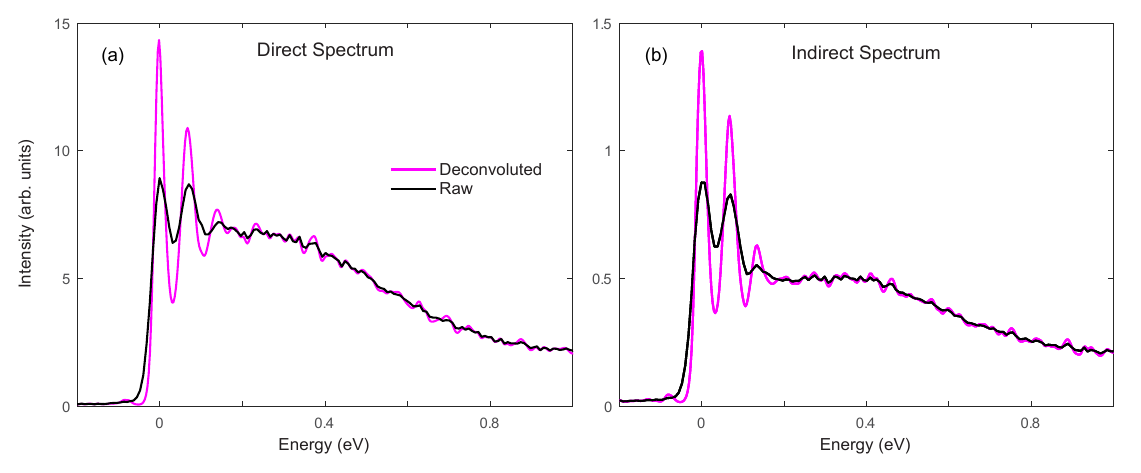}
    \caption{Deconvoluted direct and indirect spectra over laid with corresponding raw spectra from the polarimetric experiment. Deconvoluted spectra were used to calculate the spectrum Fig. 2 in main text.}
    \label{fig:deconv_specs}
\end{figure}

\pagebreak
\newpage

\noindent\textbf{IX. Fitting procedure to extract two-phonon intensity}

\begin{figure}[H]
    \centering
    \includegraphics[width=16cm]{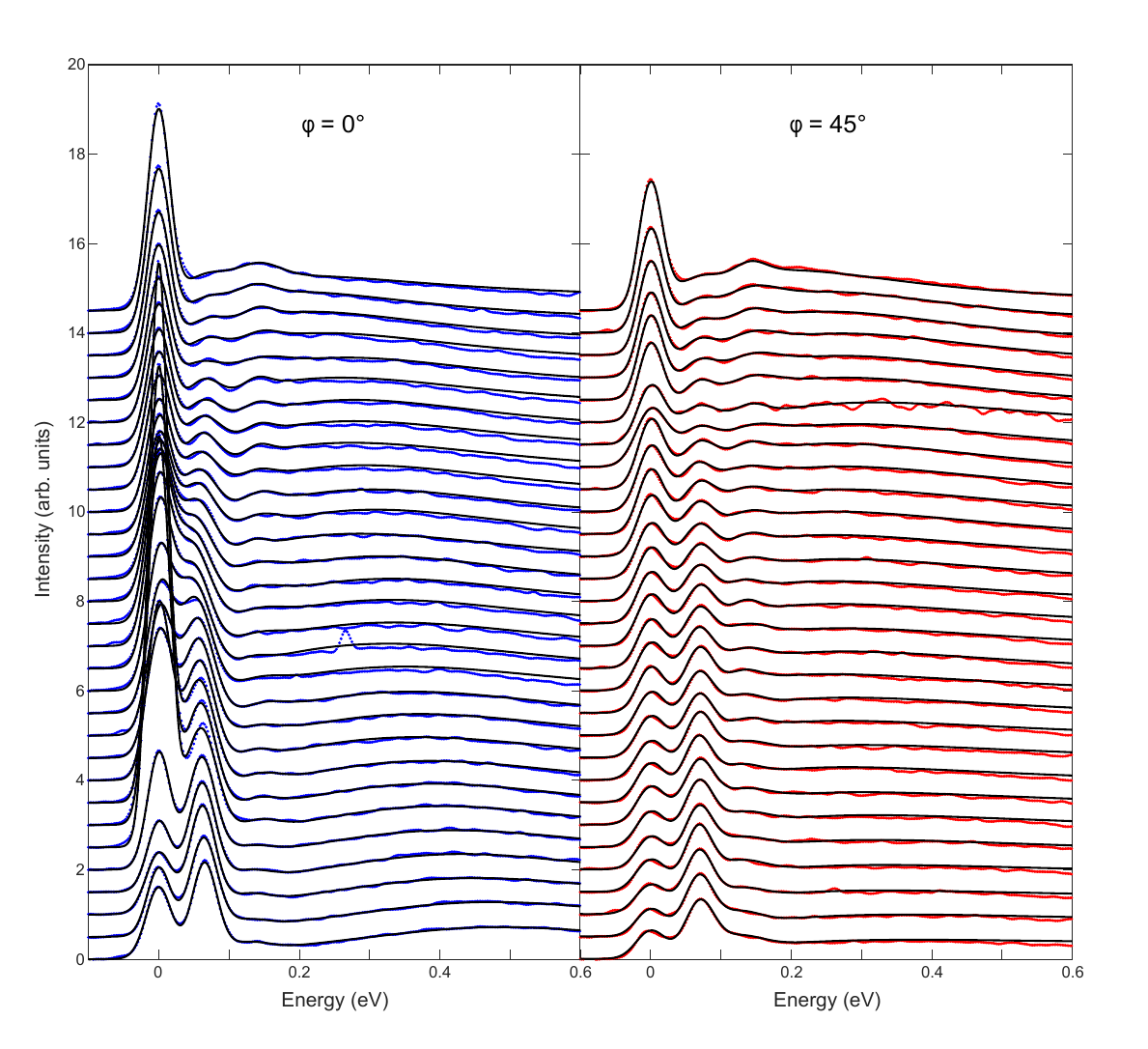}
    \caption{Comprehensive five feature fits, detailed below, to the spectra of the two high-symmetry directions for analyzing the two-phonon mode in Fig. 4 of main text. The spectra correspond (top to bottom) to $q$-values (in r.l.u.): 0.10 to 0.16 in steps of 0.02, 0.17 to 0.37 in steps of 0.01, 0.39 to 0.47 in steps of 0.02.}
    \label{fig:all highsymm fits}
\end{figure}

\bibliography{apssamp}